\documentclass[aps,prl,reprint,groupedaddress,showpacs,amsmath,twocolumn,amssymb,floatfix]{revtex4-1}
\usepackage{graphicx}
\usepackage{bm}
\usepackage{color}
\newcommand{\hatl}{$\bm{\hat{l}}$}
\newcommand{\dee}{$\bm{\hat{d}}$}
\newcommand{\he}{$^3$He}
\newcommand{\ve}{$\bm{\varepsilon}$}
\newcommand{\vfield}{$\bm{H}$}
\newcommand{\vell}{$\hat{\ell}$}
\newcommand{\aaa}{$\mathbf{a0}$}
\newcommand{\aab}{$\mathbf{a19}$}
\newcommand{\aac}{$\mathbf{a23}$}
\newcommand{\bba}{$\mathbf{b12}$}
\newcommand{\bbb}{$\mathbf{b20}$}
\newcommand{\bbc}{$\mathbf{b30}$}	
\newcommand{\elperph}{\vell\ $\perp$ \vfield}

\newcommand{\red}{\textcolor{black}}
\newcommand{\blue}{\textcolor{black}}

\begin{document}

\title{Anisotropic phases of superfluid $^3$He in compressed aerogel}

\author{J.I.A. Li}
\email[]{jiali2015@u.northwestern.edu}
\author{A.M. Zimmerman}
\author{J. Pollanen}
\thanks{Present Address: Institute for Quantum Information and Matter (IQIM), California Institute of Technology, Pasadena, California 91125, USA}
\author{C.A. Collett}
\author{W.P. Halperin}
\email[]{w-halperin@northwestern.edu}
\affiliation{Northwestern University, Evanston, IL 60208, USA}

\date{\today}

\begin{abstract}
It has been shown that the relative stabilities of various superfluid states of $^3$He can be  influenced by  anisotropy in a silica aerogel framework. We  prepared a suite of aerogel samples compressed up to 30\% for which we performed pulsed NMR on $^3$He imbibed within the aerogel. We identified A and B-phases and  determined their magnetic field-temperature phase diagrams as a function of strain.  From these results we infer that the B-phase \blue{is distorted by} negative strain \blue{forming an} anisotropic superfluid state more stable than the A-phase.

\end{abstract}
%\pacs{67.30.Hm, 67.30.Er, 67.30.Hj, 74.20.Rp}

\maketitle

The  A and B-phases of superfluid $^3$He are unconventional Cooper paired states with orbital angular momentum $L=1$.  They are just two of many $p$-wave states each distinguished by unique  broken symmetries and these are the only stable states in pure $^3$He in low magnetic field~\cite{Vol.90}. However, new phases in this manifold have been reported for superfluid confined within an anisotropic external framework; in one case chiral states within positively strained 98\% porous, silica aerogel~\cite{Pol.12a} and in another, equal spin pairing states within nematically oriented alumina hydrate aerogels~\cite{Ask.12}.  Additionally, for a negatively strained aerogel, obtained by compression, we found that the relative thermodynamic stability between the A and B-phases can be reversed resulting in an unusual tricritical point between A, B, and normal phases with a critical magnetic field of $H_c \approx 100$ mT~\cite{Li.14b}.  

Our observation of preferential B-phase stability relative to the A-phase ~\cite{Li.14b} appears at first glance to be inconsistent with theoretical predictions where it was established that anisotropic  scattering of $^3$He quasiparticles would favor the anisotropic A-phase, over the isotropic B-phase ~\cite{Thu.98, Aoy.06, Vic.05}.  To address this problem we have investigated the dependence of the A-to-B phase diagram as a function of negative strain. 

Here we report that the square of this critical field  is directly proportional to the aerogel anisotropy which we determined from optical birefringence.  Based on our measurements of the strain dependence of the A-to-B phase diagram we have determined that the B-phase entropy is increased relative to the unstrained B-phase which  transforms under negative strain to a new type of anisotropic superfluid, which we call a distorted B-phase, that can effectively compete with the A-phase in a magnetic field.  

%%%%%%%%%%%%%%%%%%%%%%%%%%%%%%%%%%%%%%%%%%%%%%%%%%%%
%%%%%%%%%%%%%%%%%%       Figure 1       %%%%%%%%%%%%%%%%%%%%%%%%%%
%%%%%%%%%%%%%%%%%%%%%%%%%%%%%%%%%%%%%%%%%%%%%%%%%%%%
\begin{figure}
\centerline{\includegraphics[height=0.40\textheight]{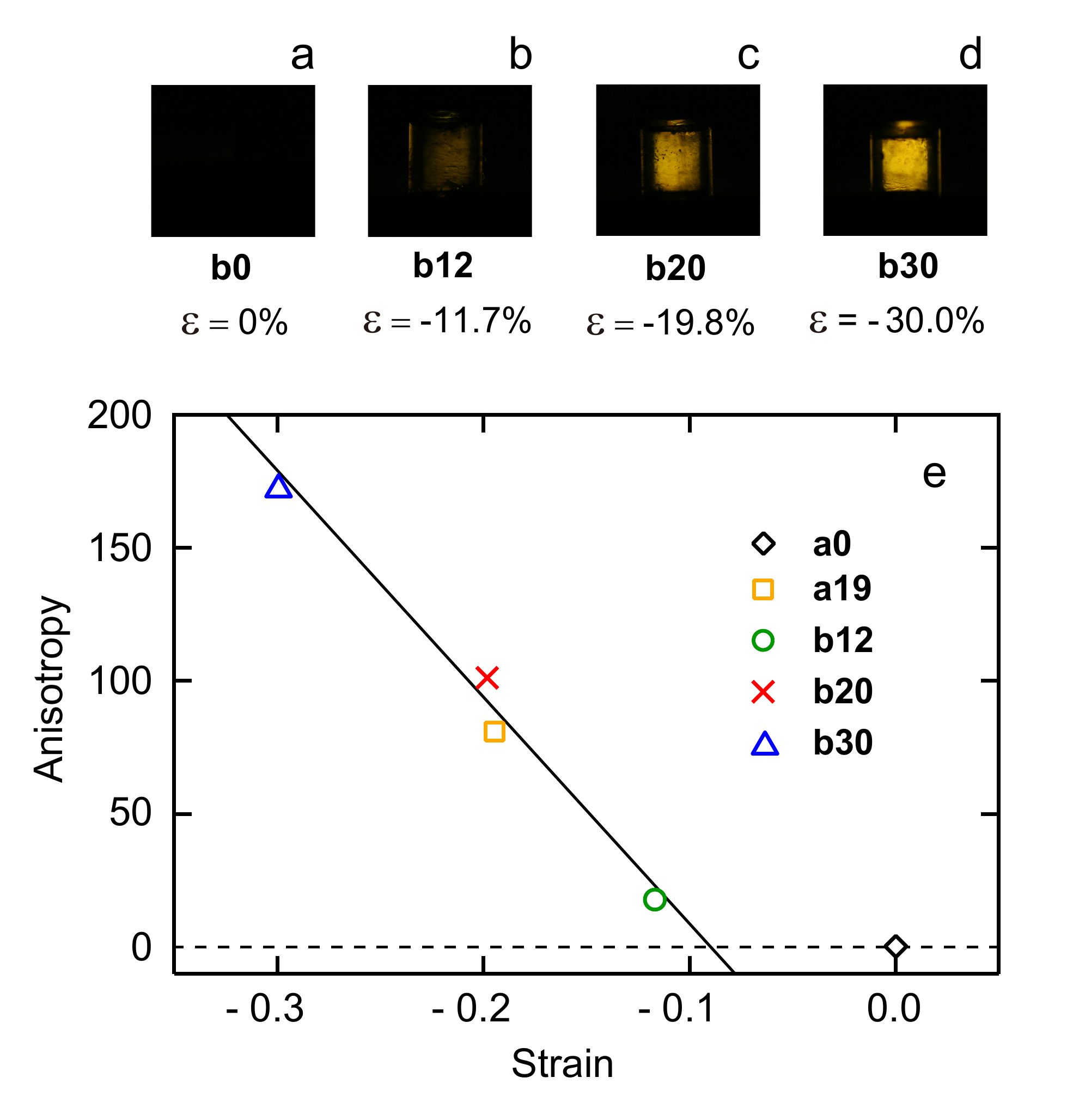}}
\caption{\label{fig1}(Color online).  Anisotropy introduced by compressive strain, \ve, in two batches of  98\% porosity aerogel, {\bf a} and {\bf b}. a) to d)  Optical birefringence images of the  transmitted light intensity through the aerogel samples.  e) Anisotropy in the aerogel defined as the intensity of the optical birefringent signal versus strain. The black solid line is a linear fit for samples from batch $\mathbf{b}$. }
\end{figure}

We prepared two different batches of aerogel samples of  $98\%$ porosity, batch $\mathbf{a}$ and $\mathbf{b}$,  in the shape of cylinders $4.0$ mm in diameter with unstrained length of $5.1$ mm. For clarity we label the sample by the batch number as well as the nominal amount of negative strain introduced by compression, \emph{e.g.}, a sample from batch $\mathbf{a}$ with $-19.4\%$  negative strain is labeled as sample $\mathbf{a19}$. Some results with samples from batch $\mathbf{a}$ have also been reported previously~\cite{Pol.11,Li.12,Li.13,Li.14b}, where the superfluid phases were identified and the amplitude of the order parameter was measured.  A critical field was discovered for samples with  negative strain $-19.4\%$ ($\mathbf{a19}$) and $-22.5\%$($\mathbf{a23}$) introduced into isotropic samples from the same batch
 ($\mathbf{a0}$); $\mathbf{a19}$ \blue{(originally {\bf a0})} was measured with the strain axis parallel to the magnetic field, \ve\,$\parallel\,${\bf H}, and $\mathbf{a23}$ with \ve\ $\perp$\,{\bf H}. In the present work three new samples are taken from batch $\mathbf{b}$  prepared using the same methods as batch $\mathbf{a}$. Optical birefringence was performed before introducing anisotropy by compression and the samples were found to be homogeneous and isotropic, Fig.~1(a)~\cite{Pol.08,Zim.13}.  We strained these samples $-11.7\%$ ($\mathbf{b12}$), $-19.8\%$ ($\mathbf{b20}$), and $-30.0\%$ ($\mathbf{b30}$) by mechanical compression measured as a displacement at room temperature. The strain induced anisotropy was characterized by the intensity from transmitted optical birefringence, Fig.~1(b-d), taken from a 2 $\times$ 2 mm$^2$ central portion of the images.  The integrated intensity as a function of strain for the two different batches are consistent with each other, Fig.~1(e), both of which show a delayed onset of birefringence with strain of \ve\ $\approx -8\%$, due to the intrinsic mechanical behavior of the aerogel (see supplementary information).  Beyond this onset, the transmitted intensity increases linearly. All  $\mathbf{b}$ samples were oriented with the strain axis parallel to the  field, cooled simultaneously, and filled with $^3$He supercritically. We performed pulsed NMR measurements at a pressure $P$\,=\,26.5 bar in magnetic fields ranging from $H$\,= \,49.1 to 214 mT with small tip angles, $\beta \sim 20^{\circ}$.   Detailed procedures for these measurements and their analyses have been described earlier~\cite{Pol.11,Pol.12a,Li.13}.

%%%%%%%%%%%%%%%%%%%%%%%%%%%%%%%%%%%%%%%%%%%%%%%%%%%%
%%%%%%%%%%%%%%%%%%       Figure 2       %%%%%%%%%%%%%%%%%%%%%%%%%%
%%%%%%%%%%%%%%%%%%%%%%%%%%%%%%%%%%%%%%%%%%%%%%%%%%%%
\begin{figure}
\centerline{\includegraphics[height=0.27\textheight]{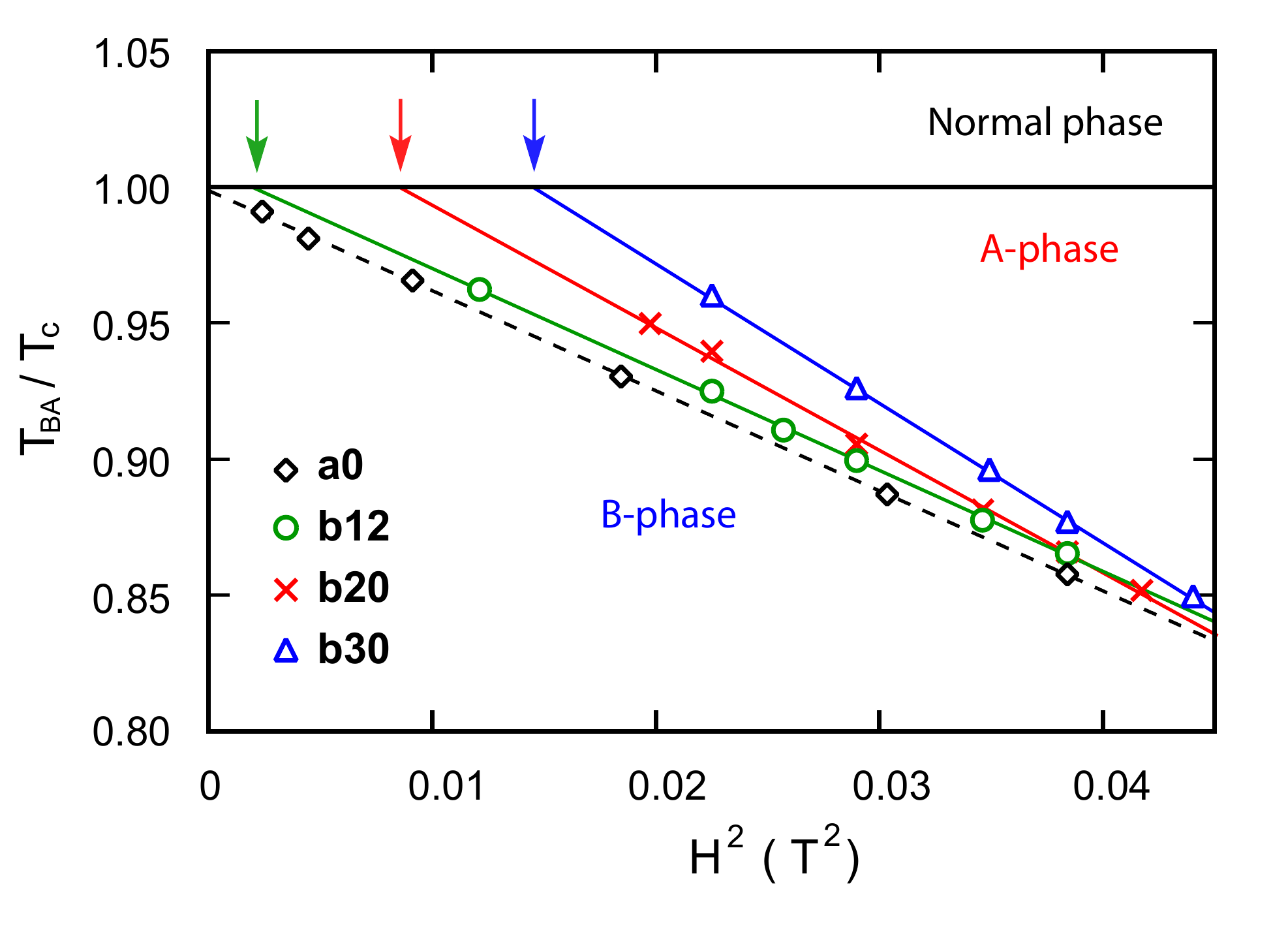}}
\caption{\label{fig2} (Color online).  Superfluid phase diagram $T_{BA}/T_c$ versus $H^2$, with \ve \,$\parallel\bm{H}$ for anisotropic $\mathbf{b}$ samples compared to the isotropic sample $\mathbf{a0}$~\cite{Li.13} at a pressure of $P = 26.5$ bar from warming experiments.  The superfluid transition temperatures, $T_c$, for $\mathbf{a0}$,  $\mathbf{b12}$, $\mathbf{b20}$, and $\mathbf{b30}$ are $2.213$, $2.152$, $2.118$, and $2.118$ mK.  The critical fields are indicated by arrows:  $H_c = 45.5$ mT, $91.8$ mT, and $120.2$ mT for these samples.}
\end{figure}

The field dependence of the thermodynamic transitions on warming, $T_{BA}(H^2)$ are shown in Fig.~2 for sample $\mathbf{a0}$, $\mathbf{b12}$, $\mathbf{b20}$, and $\mathbf{b30}$. Quadratic field dependences are observed for the field suppression of $T_{BA}$ in all samples, which can be qualitatively understood by analogy with the free energy difference between  A and B-phases for pure $^3$He in the Ginzburg-Landau limit given by,

\begin{equation}
F_B-F_N =-\frac{1}{36}\frac{\left(\alpha^2(T)+2\alpha(T)g_zH^2\right)}{\beta_{12}+\frac{1}{3}\beta_{345}} +\mathit{O}\left( H\right)^4,\label{1}\\
\end{equation}
\noindent
\begin{equation}
F_A-F_N =-\frac{1}{36}\frac{\alpha^2(T)}{\beta_{245}}.\label{2}\\
\end{equation}
\noindent
The $\beta_i$ and $g_z$ are material parameters in the Ginzburg-Landau functional~\cite{Rai.76,Ser.78} that determine the superfluid condensation energy and the symmetry of the superfluid states, and $\alpha(T) \propto 1-T/T_c$. The B-phase has a quadratic field dependent free energy compared to the normal state, whereas the A-phase free energy is field independent. From Eq.~1 and 2, the field suppression of $T_{BA}$ in the GL limit can be written as,~\cite{Ger.02,Hal.08}  

\begin{equation}
1-\frac{T_{BA}}{T_c}=g_{BA}\left( \frac{H^2-H^2_c}{H^2_0}\right) +\mathit{O}\left( \frac{H}{H_{0}}\right)^4,\label{3}\\
\end{equation}
\noindent
where $H_c = 0$  for pure $^3$He and for $^3$He in an isotropic aerogel~\cite{Pol.11}.  $H_0$ and $g_{BA}$ are  constants from the GL theory~\cite{Thu.98,Cho.07}.   However, the presence of strain in the aerogel breaks the orbital rotation symmetry of the superfluid to which we attribute the appearance of a critical field, $H_c$, that we find depends systematically on the magnitude of the strain.  In fact all  anisotropic aerogel samples with negative strain exhibit this behavior as exemplified in Fig.~2. The existence of a critical field is indicative of a new term in the relative free energy between the A and B-phase which is not captured by Eq.~1 and 2.  To determine the origin of $H_c$ we must isolate how each of the phases is affected by strain.  For this purpose we measure the longitudinal resonance frequency  of the superfluid A and B-phases, $\Omega_A$ and $\Omega_B$, which are directly related to their respective order parameter amplitudes, $\Delta_A$ and $\Delta_B$, and correspondingly to their free energy.

%%%%%%%%%%%%%%%%%%%%%%%%%%%%%%%%%%%%%%%%%%%%%%%%%%%%
%%%%%%%%%%%%%%%%%%       Figure 3       %%%%%%%%%%%%%%%%%%%%%%%%%%
%%%%%%%%%%%%%%%%%%%%%%%%%%%%%%%%%%%%%%%%%%%%%%%%%%%%
\begin{figure}
\centerline{\includegraphics[height=0.26\textheight]{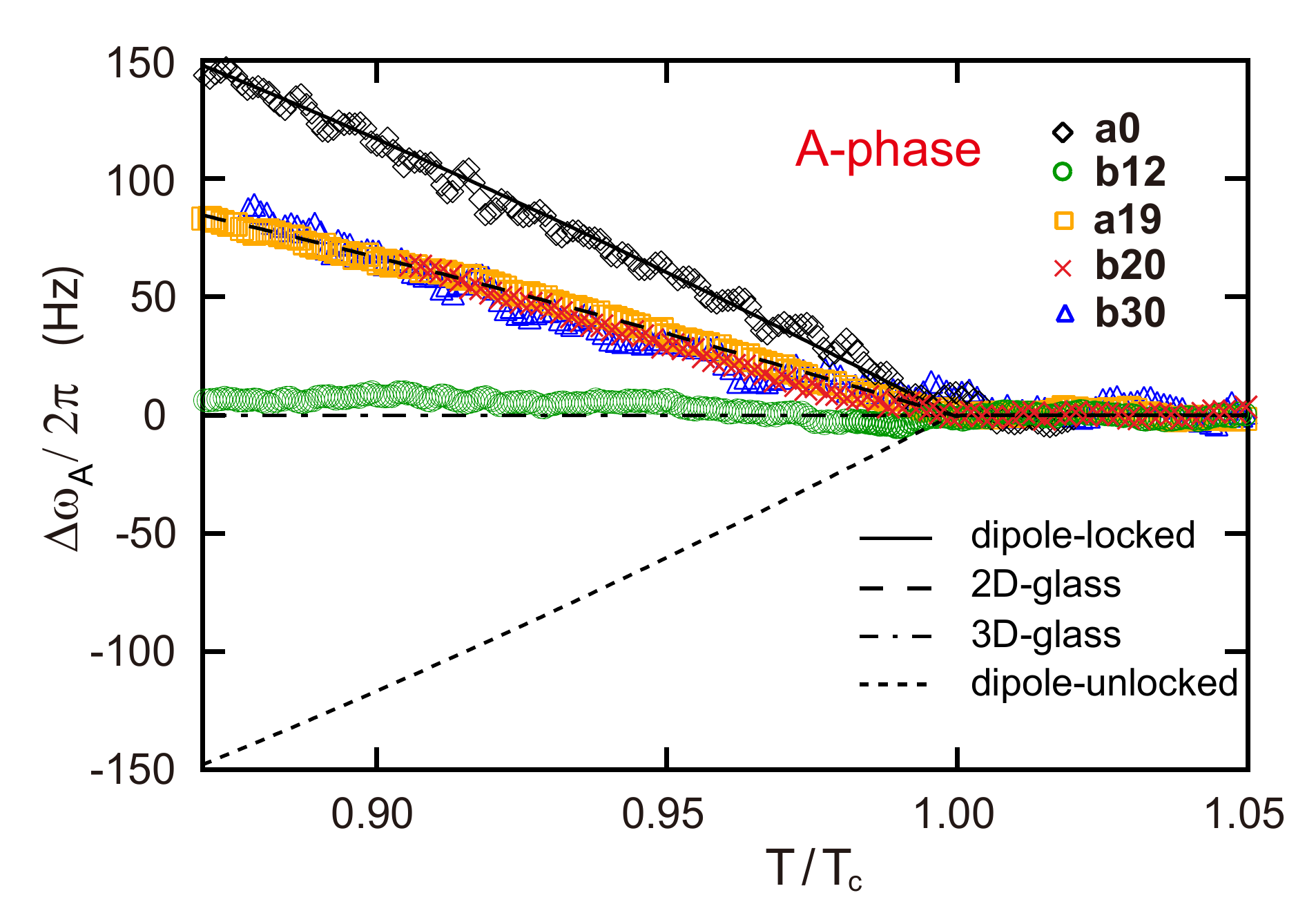}}
\caption{\label{fig3} (Color online). NMR frequency shifts of the A-phase, $\Delta\omega_{A}$,  as a function of reduced temperature measured with small tip angle, at $P \approx 26$ bar, adjusted to a common field of $H=196$ mT.  On warming from the B-phase, sample $\mathbf{a0}$ is in the dipole-locked (\dee\,$\parallel$\,\hatl\,$\perp$\,\vfield) configuration giving a maximum frequency shift (black diamonds)~\cite{Pol.11}. The black solid curve is for \red{the frequency shift of} pure \he-A with a scaling factor of \red{$0.42$, corresponding to an order parameter suppression of $0.35$}. The frequency shift of a two-dimensional glass phase at small tip angle is calculated using Eq.~5, black dashed curve, in good agreement with the data for  $\mathbf{a19}$ at $174$ mT (orange squares) and for $\mathbf{b20}$ (red crosses) and $\mathbf{b30}$  ( blue triangles) at $196$ mT.  The negligible shift for  \bba\ (green circles) with no linewidth increase (not shown), is characteristic of a three-dimensional glass phase~\cite{Li.13}.}
\end{figure}

For the isotropic sample $\mathbf{a0}$, $\Omega_A$ and $\Omega_B$ satisfy the Leggett relation, Eq.~4, which expresses the unique order parameter symmetries of the A and B-phases in the Ginzburg-Landau regime given by, 
	\begin{equation}
	\frac{5}{2}=\frac{\Omega_{B}^{2}}{\Omega_{A}^{2}}\frac{\chi_{B}}{\chi_{A}}\frac{\Delta_{A}^{2}}{\Delta_{B}^{2}}\label{4}
	\end{equation}
\noindent
where we  take  $\Delta_{A}^{2} \approx \Delta_{B}^{2}$~\cite{Pol.11,Leg.75} \footnote{the free energy difference between the two phases is much smaller than the superfluid condensation energy \cite{Bau.04}} and the susceptibilities $\chi_{A}$ and $\chi_{B}$ are measured.

The maximum frequency shift of the A-phase from sample $\mathbf{a0}$ is shown in Fig.~3 as open black diamonds, where the entire sample is in the dipole-locked configuration (\dee\,$\parallel$\,\hatl\,$\perp$\,\vfield) ensured by warming from the B-phase~\cite{Li.13}, where \dee\ is the direction along which the spin projection is zero, and \hatl\ is the direction of the orbital angular momentum.  The longitudinal resonance frequency can be calculated from the high field approximation: ${\Omega_A}^2=2\omega_L\,\Delta\omega_{A}$ with $\Delta\omega_{A}$ the measured shift in frequency from the Larmor frequency $\omega_{L}$. When our isotropic aerogel is subjected to sufficient uniaxial negative strain the A-phase becomes a two-dimensional (2D) orbital glass phase~\cite{Li.14b}  with frequency shift at small NMR tip angles $\beta$ given by ~\cite{Dmi.10,Elb.08},
\begin{equation}
\Delta\omega_{2D}= \frac{{\Omega_A}^2}{4\omega_L}\hspace{20pt}\beta \approx 0.\label{5}
\end{equation}
Using $\Omega_A$ determined from the isotropic sample $\mathbf{a0}$ and Eq.~5, the frequency shift for a 2D glass phase with the same order parameter amplitude is plotted as the black dashed curve in Fig.~3. The frequency shift for samples $\mathbf{a19}$, $\mathbf{b20}$ and $\mathbf{b30}$, Fig.~3, agree well with the black dashed curve, showing that the longitudinal resonance frequency of the A-phase, $\Omega_A$, is not affected by negative strain up to $30\%$. We conclude that the order parameter amplitude of the A-phase, $\Delta_A$, is insensitive to aerogel anisotropy induced by negative strain and consequently the free energy is also unaffected. Additionally,  a zero frequency shift and zero linewidth increase are observed in sample \bba, Fig.~3, indicative of the existence of a three-dimensional (3D) glass phase~\cite{Li.13}. As shown in Fig.~1(b) and (e), sample \bba\ is strained significantly by \ve\ $= -11.7\%$ but only displays a weak optical birefringence due to the delayed onset of anisotropy at $\sim -8\%$ strain. Therefore, the observation of a 3D glass phase suggests a threshold aerogel anisotropy below which the continuous 3D rotational symmetry of the order parameter, and thus the 3D glass phase, is preserved (see supplementary information). \blue{Consequently, the transition from a 3D-glass phase to a 2D-glass phase must occur with increasing (negative) strain within a small $7\%$ window.} The NMR tip angle behavior of the 3D glass phase in sample \bba\ is shown in the supplementary information.

%%%%%%%%%%%%%%%%%%%%%%%%%%%%%%%%%%%%%%%%%%%%%%%%%%%%
%%%%%%%%%%%%%%%%%%       Figure 4       %%%%%%%%%%%%%%%%%%%%%%%%%%
%%%%%%%%%%%%%%%%%%%%%%%%%%%%%%%%%%%%%%%%%%%%%%%%%%%%
\begin{figure}
\centerline{\includegraphics[height=0.26\textheight]{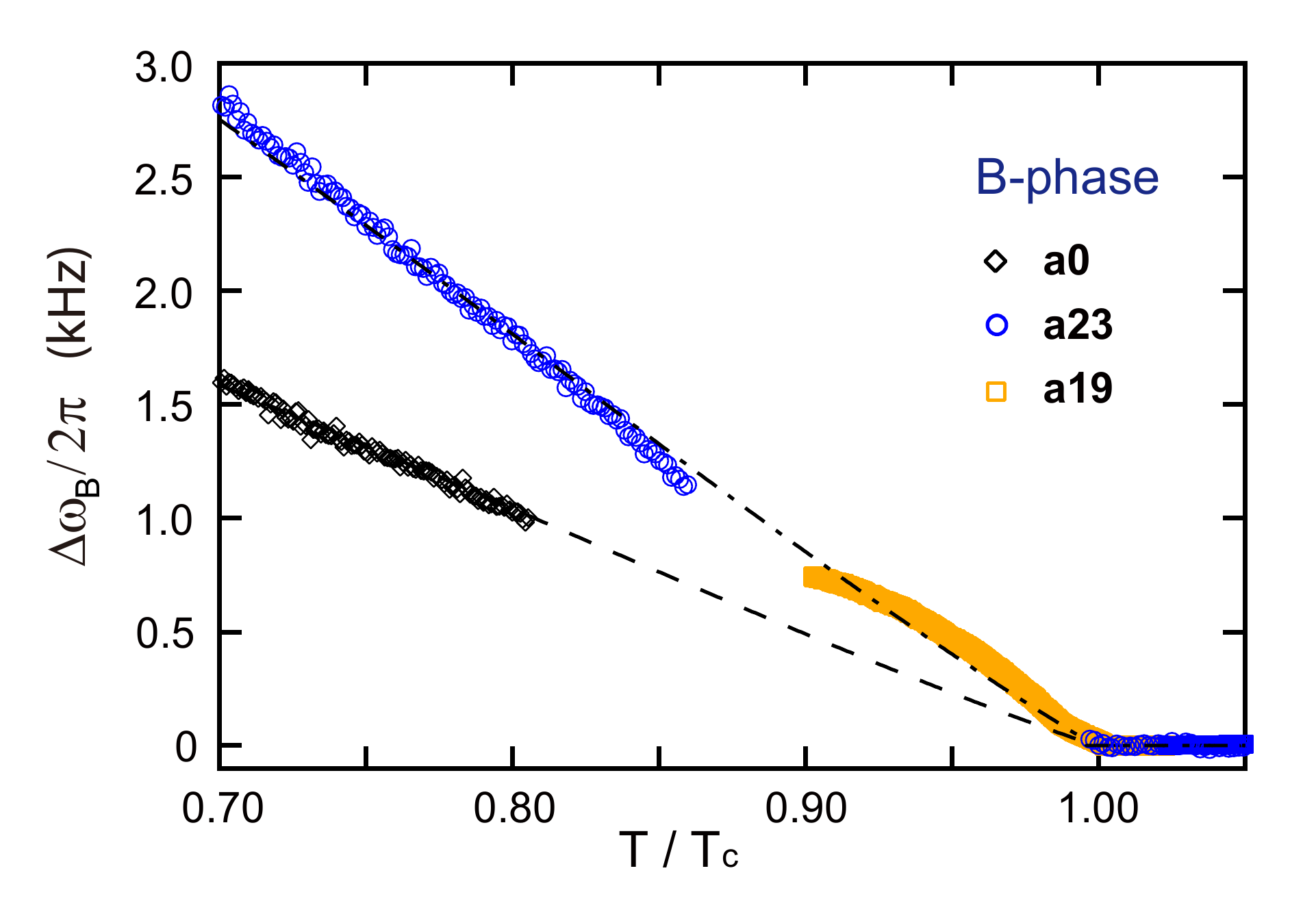}}
\caption{\label{fig4} (Color online). NMR frequency shifts $\Delta\omega_{B}$ of the B-phase as a function of reduced temperature measured with small tip angle, at $P=26.3$ bar.  The frequency shifts  are from a peak in the NMR spectrum~\cite{Pol.11} corresponding to the texture, \vell\,$\perp$\,\vfield, that determines $\Omega_B$, Eq.~\ref{6}. \blue{This texture appears above (below) a crossover temperature $T_x \sim 0.85 T_c$ for sample {\bf a19} ({\bf a23}); see supplementary information}. The frequency shifts for the anisotropic samples $\mathbf{a19}$  and $\mathbf{a23}$, measured at $H = 95.6$ mT,  are significantly larger than for the isotropic sample $\mathbf{a0}$ indicating a change in the order parameter for strain  \ve\ $\sim-20\%$. The frequency shift for sample \aaa\ was measured at $H = 196$ mT and scaled to a common field of $95.6$ mT using Eq.~\ref{6}. \blue{A comparison with samples {\bf b20} and  {\bf b30} was not possible owing to texture inhomogeneity for $T>T_x$}.}
\end{figure}

In the B-phase the order parameter texture with orbital quantization axis \vell\ perpendicular to the magnetic field, for example constrained by  sample walls,  has a  maximum frequency shift for small $\beta$  directly related to the longitudinal resonance frequency and can be  identified as a prominent peak in the spectrum at  large frequency ~\cite{Pol.11}, 
			\begin{eqnarray}
      \Delta\omega_{B}=\frac{2}{5}\frac{\Omega_{B}^{2}}{\omega_{L}}\hspace{20pt}\beta \approx 0.\label{6}
      \end{eqnarray}

\noindent
We refer to the spin dynamics of this texture as the \vell\,$\perp$\,\vfield-mode.  In previous work we measured  $\Delta\omega_{B}$, for this mode in sample $\mathbf{a0}$~\cite{Pol.11}, shown in Fig.~4 together with its extension to $T_c$ as a black dashed curve.  When we compare this to the frequency shift in the A-phase, Fig.~3, we obtain  the factor of 5/2 given in Eq.~4; this factor expresses the characteristic symmetries of the axial and isotropic $p$-wave states (A and B-phases respectively)  as is the case for pure superfluid $^3$He.  On the other hand, for the negatively strained aerogels  $\mathbf{a19}$ and $\mathbf{a23}$, the \vell\ $\perp$\,\vfield-mode frequency shifts   are significantly increased.  The black dash-dotted curve is a fit to our data for $H=95.6$ mT found by multiplying the black dashed curve by $1.7$.  We conclude that the B-phase in strained aerogel is not an isotropic state and might have a different order parameter symmetry.  This is a new type of anisotropic superfluid which continuously evolves from the isotropic B-phase with increasing anisotropy induced by the compressive strain in the aerogel.   	

In Eq.~3, the two parameters affected by strain deduced from the AB-phase diagram are the square of the critical field, $H^2_c$,  and the slope of the B to A-phase transition, $g_{BA}/H^2_0$. In Fig.~5, we plot these two quantities as a function of anisotropy for the three aerogel $\mathbf{b}$ samples compared with the isotropic sample $\mathbf{a0}$ and find that they are proportional to the anisotropy induced by strain.  We understand that the mechanism by which strain modifies the superfluid is to create anisotropic quasiparticle scattering in the superfluid. Consequently we can infer that anisotropic scattering smoothly increases with aerogel anisotropy and is responsible for creating a distorted B-phase. Although we have not made a specific identification of the new state it is nonetheless clear from our susceptibility measurements that it is a non-equal spin pairing state (NESP) with the same susceptibility as the B-phase. The latter comparison was made between \aab\  and in the unstrained aerogel $\mathbf{a0}$ at the same temperature, pressure, and field on their respective AB-phase boundaries ~\cite{Li.14c} \red{following the procedure described in} Ref. ~\cite{Pol.11,Li.13,Li.14b}. 

%%%%%%%%%%%%%%%%%%%%%%%%%%%%%%%%%%%%%%%%%%%%%%%%%%%%
%%%%%%%%%%%%%%%%%%       Figure 5       %%%%%%%%%%%%%%%%%%%%%%%%%%
%%%%%%%%%%%%%%%%%%%%%%%%%%%%%%%%%%%%%%%%%%%%%%%%%%%%
\begin{figure}
\centerline{\includegraphics[height=0.4\textheight]{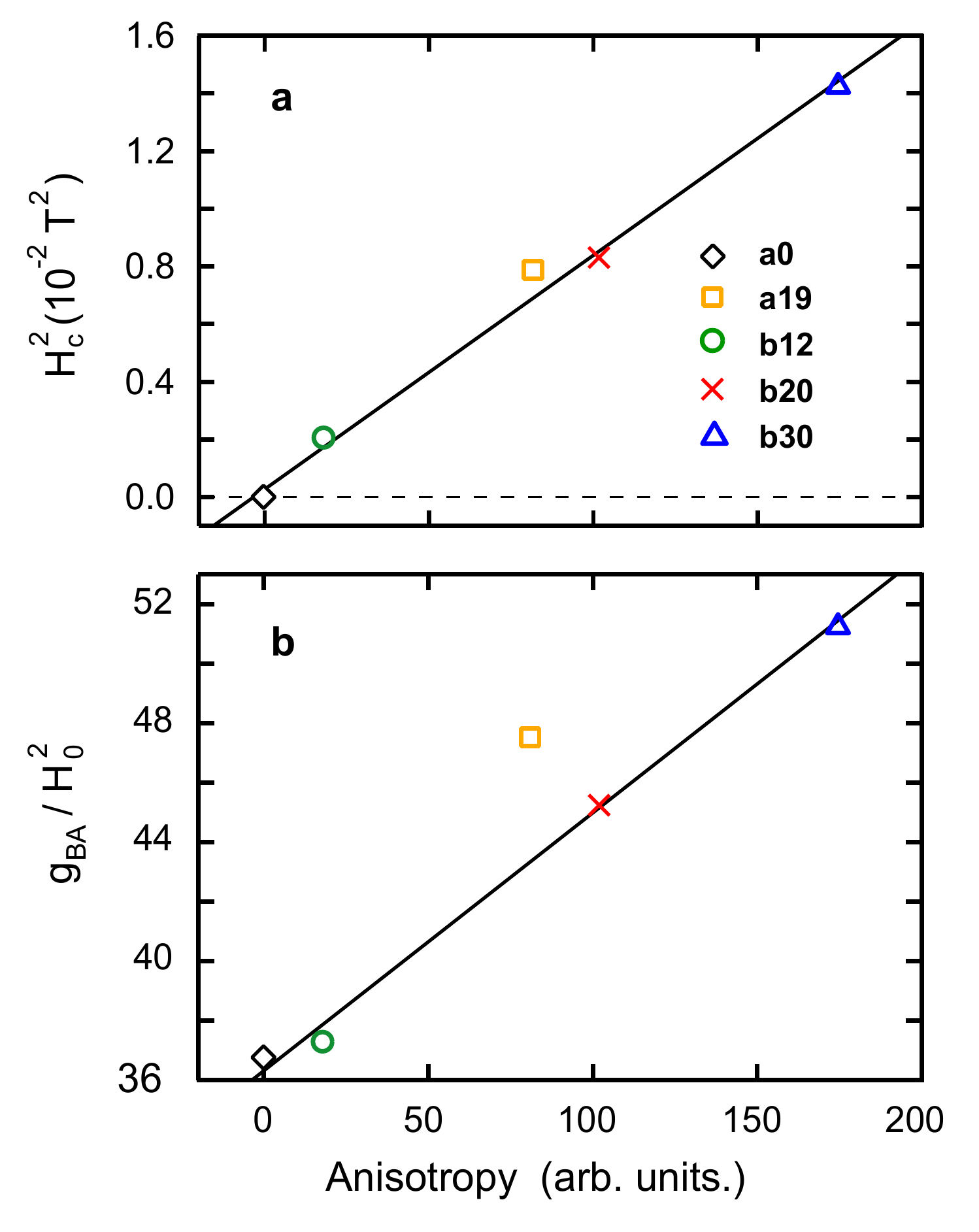}}
\caption{\label{fig5} (Color online). (a) The critical field squared $H_{c}^{2}$ and (b) $g_{BA}/H^2_0$ for sample $\mathbf{a0}$ \red{(open black diamond)}, $\mathbf{a19}$ \red{(open orange square)}, $\mathbf{b12}$ (open green circles), $\mathbf{b20}$ \red{(red crosses)} and $\mathbf{b30}$ (open blue triangles) as a function of aerogel anisotropy taken from measurements of the optical birefringence intensity. ${H_c}^2$, and  $g_{BA}/H^2_0$ show a strict proportionality to aerogel anisotropy, with the exception of sample \aab. Sample \aab\ does not lie on the same line as the batch $\mathbf{b}$ samples for the second parameter, $g_{BA}/H^2_0$. }
\end{figure}

With this input we can calculate the change of entropy of the new phase, compared to the isotropic B-phase, using the Clausius-Clapeyron equation, 
\begin{equation}
\frac{\mathrm{d}T}{\,\,\,\,\mathrm{d}H^2} = - \frac{1}{2}\frac{\chi_A - \chi_{new}}{S_A - S_{new}},\label{7}
\end{equation}
 
 \noindent
where $\chi$ and $S$ are the susceptibility and the entropy of the two phases along the phase equilibrium line.  Since the slopes of these lines, Fig.~2, are increasingly more negative with increasing strain we deduce that the new phase has a higher entropy compared with the isotropic aerogel B-phase.

In summary, we report NMR measurements as a function of compressive strain in aerogel that determine  the phase diagrams of the superfluid phases within the aerogel framework. The anisotropy in the aerogel stabilizes a new anisotropic non-equal spin pairing superfluid state which evolves continuously from the isotropic B-phase with increasing strain.  The tricritical point between normal $^3$He, the A-phase, and the distorted B-phase is marked by a critical field for which the square of the field  is proportional to the anisotropy.  Additionally,  the entropy of this distorted B-phase is larger than that of the isotropic B-phase. 

\begin{acknowledgments}
We are grateful to J.J. Wiman, J.A. Sauls, V.V. Dmitriev, Yoonseok Lee, J.M. Parpia, and G.E. Volovik for helpful discussion and to W.J. Gannon for help in the early phases of this work and to the National Science Foundation for support, DMR-1103625.
\end{acknowledgments}

\section{Supplementary Information}

{\bf SI.1\,\,\, Stress-strain characterization}\\

In previous work at Northwestern, Ref.~\cite{Pol.08,Pol.12b}, it has been shown that optical cross-polarization studies provide good measures of aerogel anisotropy. The absence of any optical axis was taken as evidence of a homogeneous, globally isotropic aerogel. Such a sample remains dark in the optical birefringence measurement, as shown in Fig.~1(a). As mechanical strain is introduced into the nominally isotropic sample, a well defined optical axis develops and the sample lights up as shown in Fig.~1(b)-(d). Therefore aerogel anisotropy can be defined as the intensity of the optical birefringent signal.

\begin{figure}
\centerline{\includegraphics[height=0.4\textheight]{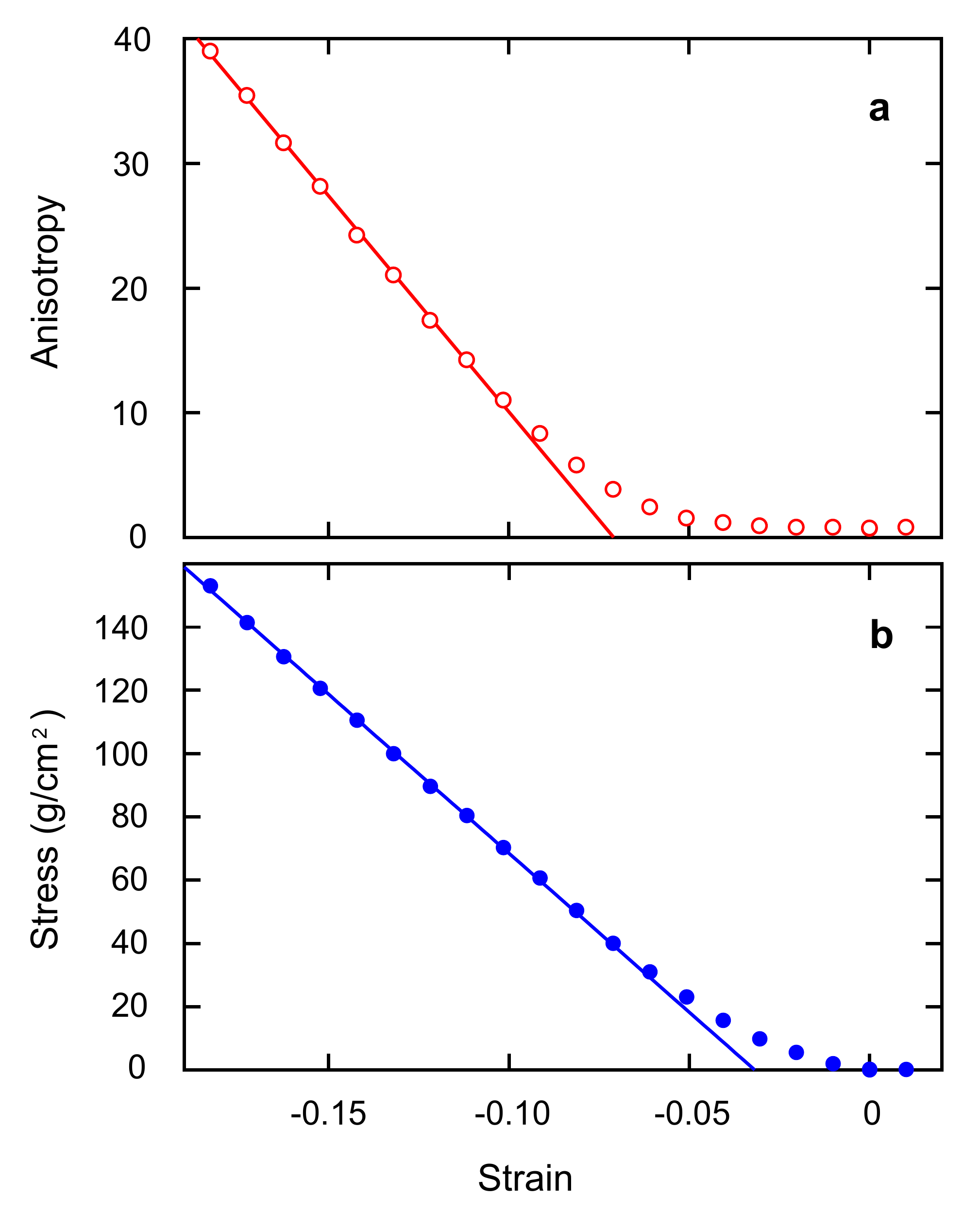}}
\caption{\label{figSI3} (Color online). Optical birefringence characterization and simultaneous stress-strain measurement of an aerogel sample from batch $\mathbf{b}$ during compression up to $\approx-20\%$. The aerogel sample has the same dimension as the samples used in our  \he\ NMR experiments. (a) Aerogel anisotropy defined as the intensity of the optical birefringence is shown versus negative strain. There is a delayed onset of anisotropy with increasing strain at \ve\ $\sim-7\%$.  (b) Stress versus strain during compression. There is a delayed onset at \ve\ $\sim-3\%$ that we attribute to a surface effect, after which the stress increases linearly with strain.  }
\end{figure}

Since the global anisotropy was introduced with mechanical strain, simultaneous stress-strain measurements were performed along with the optical characterization, on isotropic aerogel samples with different growth conditions, to determine the relationship between stress, strain and the transmitted light intensity, Ref.~\cite{Zim.13}. The aerogel samples used in Ref.~\cite{Zim.13} have cylindrical shape, $\sim 7.5$ mm in diameter and $\sim 7.5$ mm long, and the growth condition is controlled by varying ammounts of ammonia catalyst. It was discoverd that independent of the growth condition, the stress and strain response obey a linear relationship for all the samples, while the transmitted intensity versus strain displays a delayed onset at small negative strain, $\sim -3\%$. The sound velocity in the aerogel sample, calculated from the slope of the stress-strain curve, was shown to be rather sensitive to the catalyst concentration during the sample growth process, Ref.~\cite{Zim.13}, suggesting that the microscopic structure of the isotropic aerogel samples from different batches can be significantly different, due to varying growth condition. Therefore sample characterization is important before we make comparison between samples from batch $\mathbf{a}$ and $\mathbf{b}$.

\begin{figure}
\centerline{\includegraphics[height=0.23\textheight]{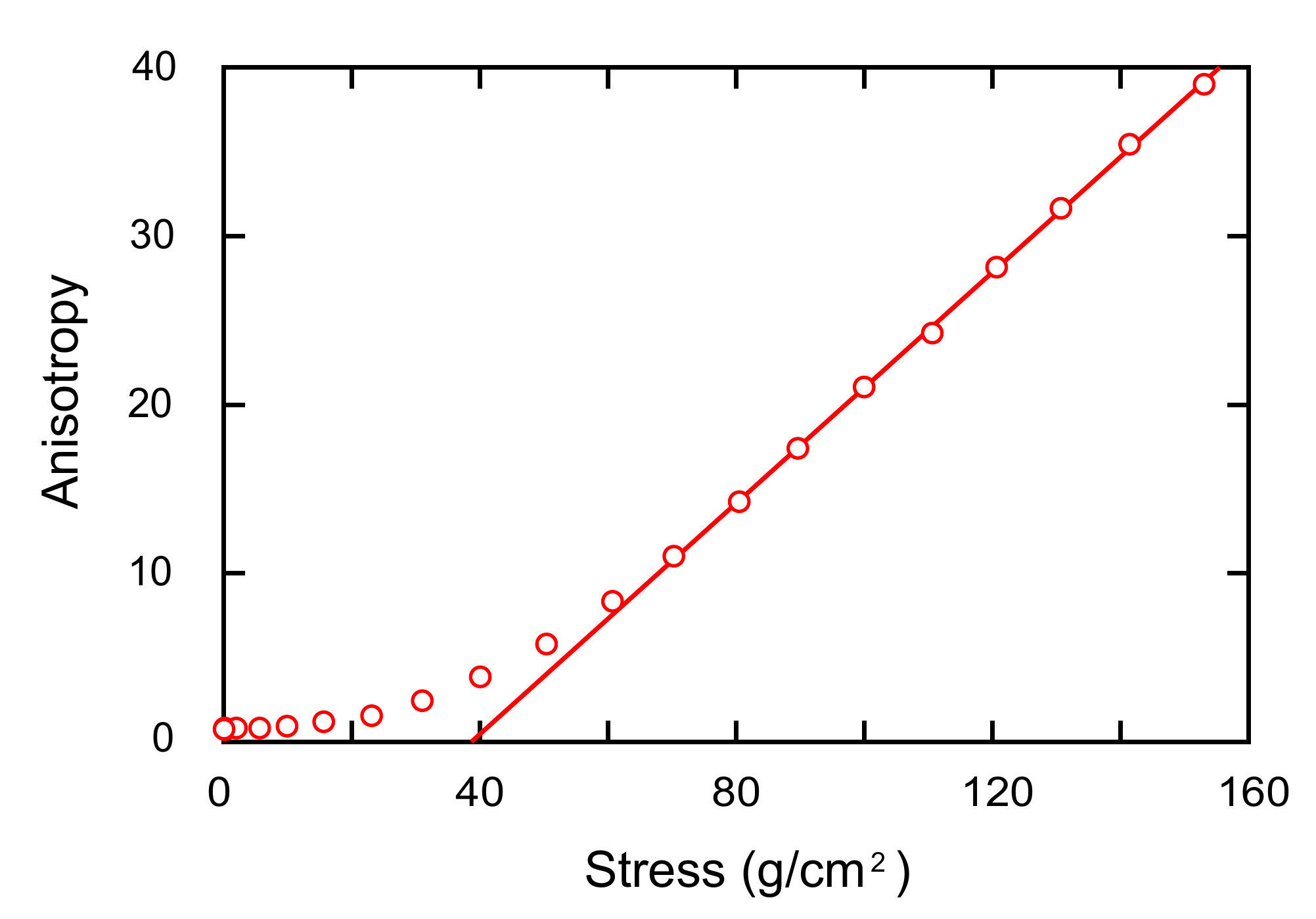}}
\caption{\label{figSI5} (Color online). Transmitted intensity versus applied stress for a batch $\mathbf{b}$ aerogel sample.  The threshold stress of $\sim40$ g/cm$^2$ is similar to our observations from other sample batches, Ref.~\cite{Zim.13}.}
\end{figure}

We performed optical birefringence characterization on an isotropic aerogel sample from batch $\mathbf{b}$ with the same dimensions as the samples used in the experiment, $4.0$ mm in diameter and $5.1$ mm long. Simultaneous stress-strain measurement, Fig.~SI\ref{figSI3}(b), onsets as negative strain is applied to the sample, providing accurate measurement for the zero of strain.  With zero of strain accurately determined, aerogel anisotropy shows a strain dependence similar to Fig.~1(e), which is linear with increased strain for negative strain of \ve\ $< -10\%$. A fit to the linear part of the birefringence intensity curve, Fig.~SI\ref{figSI3}(a), intercepts with zero intensity at a delayed onset of $\sim-7\%$ strain.

The stress and strain response shown in Fig.~SI\ref{figSI3}(b) exhibits a delayed onset %around
 \ve\ $ \lesssim -3\%$. A possible reason for this nonlinearity can be related to the fact that the surface of the aerogel sample has a different stress-strain response compared to the bulk of aerogel. This surface effect is less significant for the larger samples used in Ref.~\cite{Zim.13}. We note that the surface effect can be eliminated by plotting aerogel anisotropy versus the applied stress.  For the batch $\mathbf{b}$ sample, the transmitted light intensity as a function of stress, Fig.~SI\ref{figSI5},  displays a non-linear behavior at small stress $\sim40$ g/cm$^2$ similar to the longer samples used in Ref.~\cite{Zim.13}.

		\begin{table}
  	\begin{tabular}{| c | c | c | c | c | c |}
  	\hline
    Sample & Batch & Porosity & Strain \ve & Orientation & $T_{ca}$ (mK) \\ \hline
    \aaa   & $\mathbf{a}$ & $98.2\%$ & $0$          &    -    & $2.213$                  \\ %\hline
    \aab   & $\mathbf{a}$ & $97.8\%$ & $-19.4\%$    &  \ve\ $\parallel$ \vfield & $2.194$ \\ %\hline
    \aac   & $\mathbf{a}$ & $97.7\%$ & $-22.5\%$    &  \ve\ $\perp$ \vfield & $2.186$ \\%\hline
    \bba   & $\mathbf{b}$ & $98.0\%$ & $-11.7\%$    &  \ve\ $\parallel$ \vfield & $2.152$ \\%\hline
		\bbb   & $\mathbf{b}$ & $97.8\%$ & $-19.8\%$    &  \ve\ $\parallel$ \vfield & $2.118$ \\%\hline
		\bbc   & $\mathbf{b}$ & $97.4\%$ & $-30.0\%$    &  \ve\ $\parallel$ \vfield & $2.118$ \\ \hline
		%\cca  & $\mathbf{c}$ & $97.5\%$ & $+14.3\%$    &  \ve $\parallel$ \vfield\\ \hline
  	\end{tabular}
  	\vspace{4mm}
  	\caption[Samples used in the experiments]{\label{table1} Samples used in the experiments. }
		\end{table}

% In previous optical birefringence measurements, a universal delayed onset of birefringence was observed at $\sim-3\%$ strain in cylindrical aerogel samples $\sim7.5$ mm in diameter and $\sim7.5$ mm long ~\cite{Zim.13}, inconsistent with the delayed onset of $\sim-8\%$ shown in Fig.~1(e).

In the anisotropy versus strain relation shown in Fig.~1(e), sample \aab\ agrees well with the fit to the batch $\mathbf{b}$ samples, providing %strong 
%Leo: this might be too aggressive since we do not yet have an understanding of the behavior of a19 in Fig. 5(b).  So I also deleted the last part of your last sentence below.
evidence that samples from batch $\mathbf{a}$ and $\mathbf{b}$ were grown under similar conditions. This justifies the comparison between samples from batch $\mathbf{a}$ and $\mathbf{b}$. In Table.~\ref{table1}, we list  \blue{detailed} information of the samples from batch $\mathbf{a}$ and $\mathbf{b}$.\\
%, and is consistent with our conclusion from Fig.~5(a), that the superfluid phase diagram has similar strain dependence in samples from batch $\mathbf{a}$ and $\mathbf{b}$. \\

%While the birefringence chracterization with simultaneous stress-strain measurement is consistent with Fig.~1(e), it offers a qualitative explanation for the discrepancy between the different amount of delayed onset of birefringence. In Ref~\cite{Zim.13}, the samples with bigger dimensions are less sensitive to surface effect. In fact, we find that the surface effect can be eliminated by subtracting the delayed onset in the stress-strain measurement, which is ascribed to the surface effect, from the delayed onset in the birefringence. The resulting \ve\ $\sim-4\%$ of delayed onset of birefringence is consistent with previous experiment within error.\\

\noindent
{\bf SI.2\,\,\, Superfluid glass in the A-phase of \bba}\\

\begin{figure}
\centerline{\includegraphics[height=0.23\textheight]{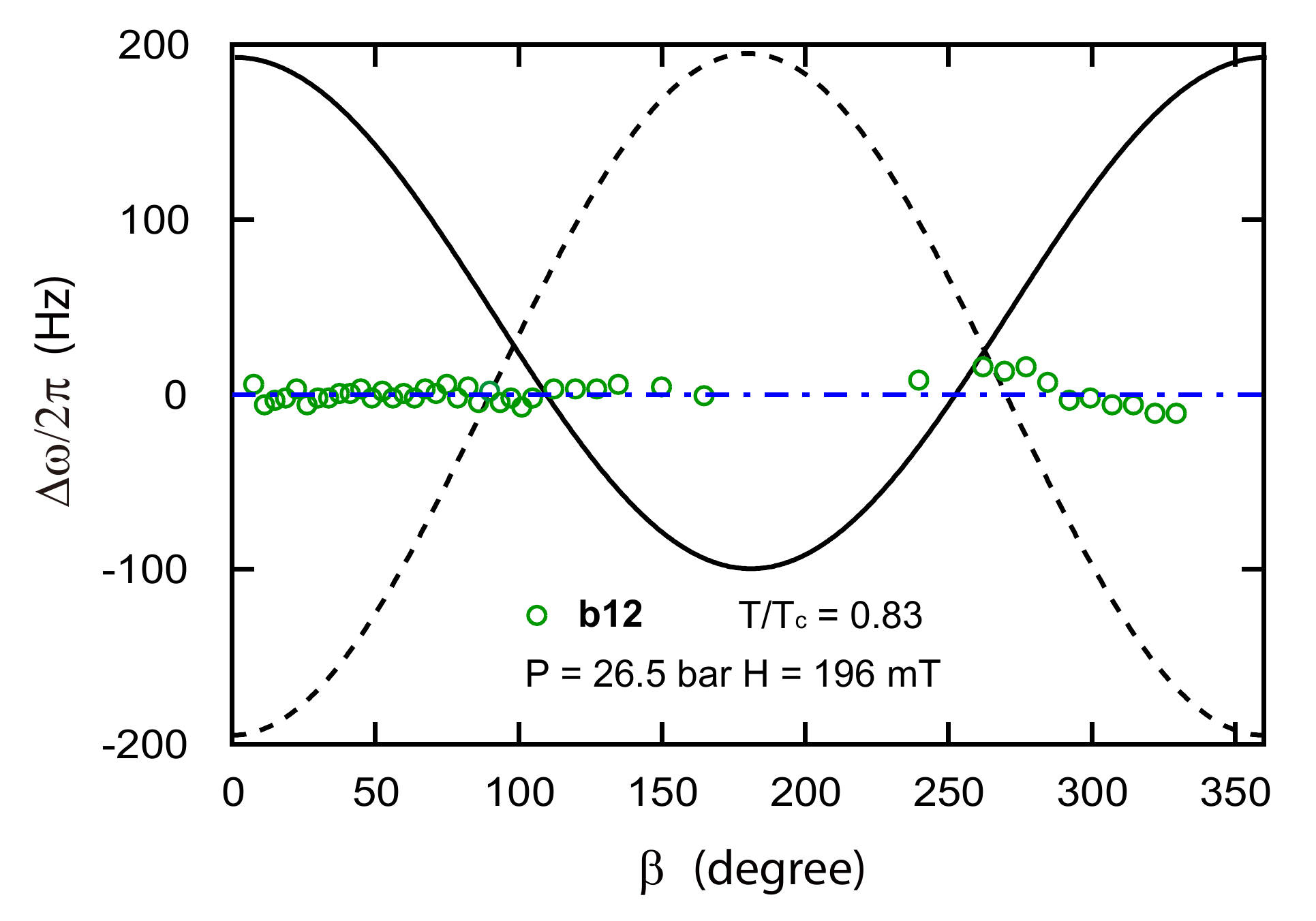}}
\caption{\label{figSI4} (Color online). Tip angle behavior of the A-phase on cooling from the normal state in sample \bba\ at $P =26.5$ bar, $H = 196$ mT and $T/T_{ca} = 0.83$.  The blue dash-dotted line is the theory for the 3D glass phase, where the superfluid order is completely hidden, Ref.~\cite{Vol.96,Li.13}, as indicated by the data. The expected tip angle behavior for the dipole-locked configuration, \hatl\,$\perp$\,\vfield, is shown by the black solid curve, and the black dashed curve is  the  dipole-unlocked configuration, \hatl \,$\parallel$\, \vfield, with $\Omega_A$ taken from the measurements for sample \aaa, Ref.~\cite{Pol.11}.   }
\end{figure}

It was predicted that the orientational order of the \he\ A-phase in aerogel would be destroyed by random fluctuations of  local anisotropy on a spatial scale less than the dipole length, $\xi_{D} \sim 10 \mu$m, thereby forming a superfluid glass, Ref.~\cite{Vol.96}. The glass phase is fundamentally different from a disordered A-phase induced by  inhomogeneity of the aerogel  on long length scales.  In order to reveal the characteristics of a  three dimensional (3D) superfluid glass it is necessary to eliminate  structural inhomogeneity requiring careful sample preparation and characterization as was demonstrated in Ref.~\cite{Li.13}.  In that work the glass phase of \he-A  was investigated in the isotropic sample  \aaa\ cooled from the normal state using NMR measurements with small tip angle $\beta$. The superfluid glass  was identified by zero frequency shift and zero linewidth broadening as compared to the normal state, consistent with theoretical prediction Ref.~\cite{Vol.96}.

The key conceptual basis of an orientational  glass phase, according to the original theoretical works of Larkin, Ref.~\cite{Lar.70}, and Imry and Ma, Ref.~\cite{Imr.75}, is the existence of a continuous symmetry of a vector order parameter that leads to the destruction of the superfluid order  on a local scale, a situation that was first noted by Volovik, Ref.~\cite{Vol.08}, to be applicable to $^3$He in aerogel.  We reported earlier, Ref.~\cite{Li.14b}, that  negative strain introduced by compression $\sim 20\%$ restores the component of the superfluid order in the direction along the strain axis, stabilizing a 2D glass phase. However, in the present work on sample \bba\ with \ve\ $= -11.7\%$ we found  almost no frequency shift, Fig.~\ref{figSI4}, and no linewidth broadening compared to the normal state.  This result  strongly indicates the formation of a 3D glass phase even in the presence of well-characterized global anisotropy, albeit quite small as indicated in Fig.~1(e).   We performed tip angle measurements of the frequency shift in sample \bba\ at $T/T_{ca} = 0.83$, deep in the superfluid A-phase. The absence of a frequency shift evident  in Fig.~SI3 for all $\beta$ up to $ \sim 360^{\circ}$, matches well with the theory for a 3D glass phase (dash-dotted blue line). The expected tip angle behavior for the dipole-locked configuration, \hatl\,$\perp$\,\vfield, is shown by the black solid curve, and the black dashed curve is  the  dipole-unlocked configuration, \hatl \,$\parallel$\vfield, with $\Omega_A$ taken from the results for sample \aaa, Ref.~\cite{Pol.11}. This is the first measurement of the NMR tip angle response of the frequency shift in a 3D superfluid glass.

Despite some global anisotropy in this sample, the observation of the 3D glass phase  indicates that the local random field anisotropy is dominant and that there exists a threshold in global anisotropy required to stabilize a 2D glass. Since a 2D glass phase is observed in sample \bbb\ and \bbc, Fig.~3, the threshold global anisotropy is between \ve\ $= -12\%$ and $20\%$, including the delayed onset in anisotropy of $\sim -8\%$.\\

%The threshold aerogel anisotropy provides evidence for the competition between the global anisotropy and the local random field anisotropy in orienting the A-phase order parameter, in the case of sample \bba, the global anisotropy induced by $12\%$ of negative strain is not enough to overcome the random field anisotropy energy.

\noindent
{\bf SI.3\,\,\, Cross-over transition in the distorted B-phase}\\

A textural crossover is ubiquitous in the B-phase for  all aerogels we have studied.  We plot the frequency shift of sample \aab\ and \aac\ versus temperature in Fig.~SI4. The textural cross-over is marked by the \blue{abrupt change} in the frequency shift data as a function of temperature.  For   all anisotropic samples the sharply defined crossover temperature is $T_{x}/T_{c} \sim 0.85$, independent of the relative orientation between the strain and the magnetic field. On warming through the transition, the orientation of the orbital angular momentum quantization axis switches from  \vell\ $\parallel$ \ve\  to  \vell\  $\perp$ \ve.   In sample \aab, the NMR spectra are dominated by the Brinkman-Smith-mode below the cross-over temperature, $T_{x} \sim 1.9$ mK, and above the transition temperature the texture shifts to  \elperph, and the Brinkman-Smith-mode is the main texture at high temperature. Due to the cross-over transition, the frequency shift of the \elperph-mode texture of the distorted B-phase can only be measured for $T < T_{x}$ for sample \aac, and above the transition temperature, $T_{x} < T < T_{ca}$ for sample \aab\ as shown in Fig.~4. 

%For $T_{x} < T < T_{c}$, the NMR spectrum of sample \aab\ is dominated by the \vell\ $\perp$ \vfield-mode as shown in Fig.~4, with frequency shift at small $\beta$ significantly increased compared to the isotropic aerogel, \aaa. \\

\begin{figure}
\centerline{\includegraphics[height=0.23\textheight]{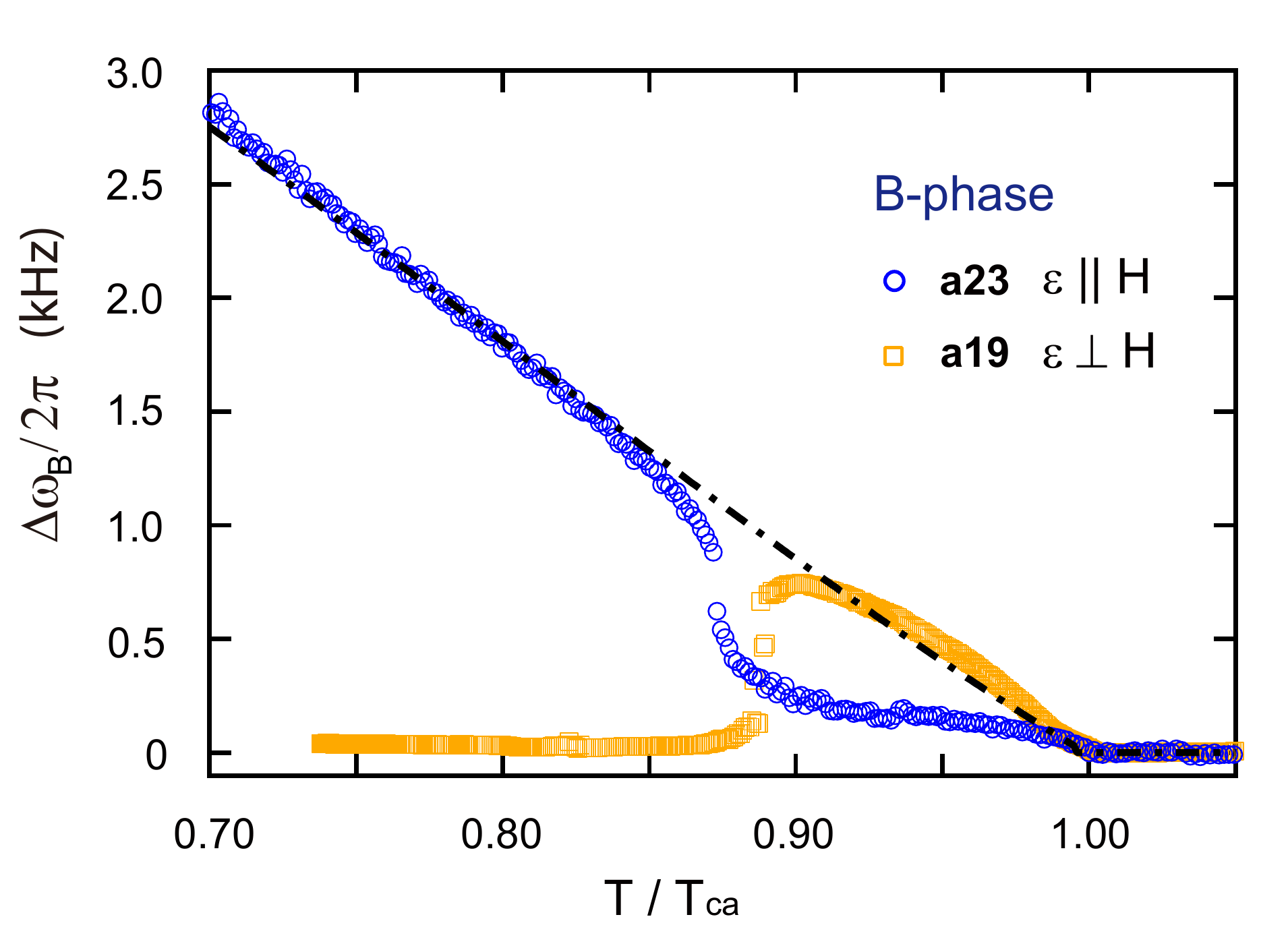}}
\caption{\label{figSI6} (Color online). B-phase frequency shift at small $\beta$, $\Delta\omega_{B}$, versus reduced temperature at $P=26.3$ bar, and $H=95.6$ mT. The cross-over transition is indicated by the sharp variation in the frequency shift in both samples. }
\end{figure}

\begin{figure}
\centerline{\includegraphics[height=0.23\textheight]{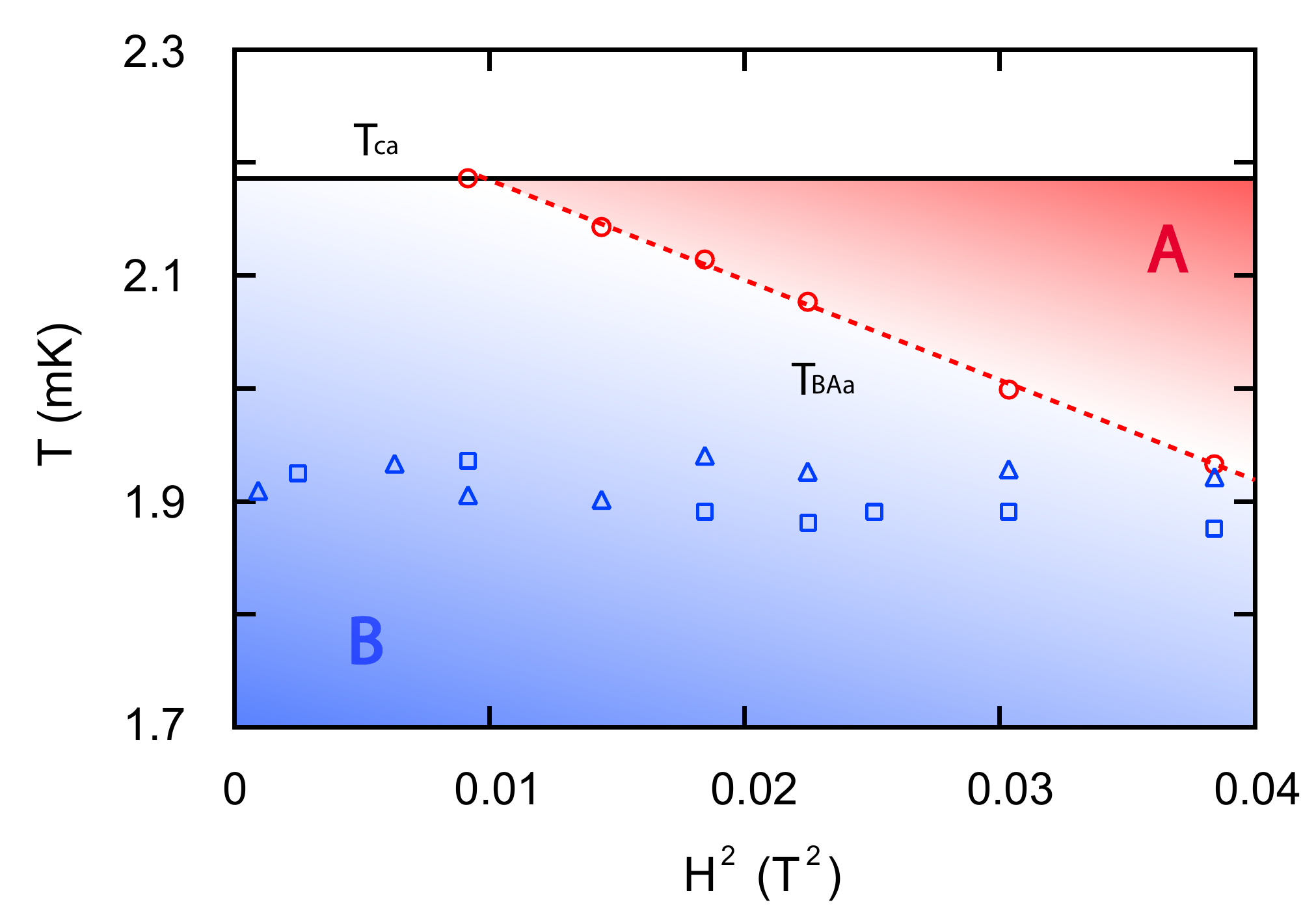}}
\caption{\label{figSI7} (Color online). Superfluid phase diagram $T/T_{ca}(H^2)$ at $P = 26.3$ bar, showing the field dependence of the cross-over transition temperature in the anisotropic samples. The black solid line is the superfluid transition in the aerogel, $T_{ca}$. The open red circles correspond to B to A-phase transitions on warming, $T_{BAa}$, measured in sample \aac. The textural transitions in the B-phase, $T_x$, for sample \aab\ (sample \aac) are shown as the blue open squares (triangles). No field dependence for $T_x$ is observed in \blue{either} anisotropic sample, and no dependence on the orientation of the field relative to strain. }
\end{figure}

 The orbital quantization axis of the B-phase can be influenced by the magnetic field, the wall of the sample cell, as well as aerogel anisotropy. Consequently a possible explanation for the textural cross-over transition is the competition between these three orienting effects. In Fig.~\ref{figSI7} we show the field dependence of the cross-over transition temperature, $T_{x}(H^2)$, in samples \aab\ and \aac. In the magnetic field range of $31.1$ to $196$ mT, the cross-over transition temperatures show no obvious field dependence, suggesting that the effect of varying the magnetic field is not significant in orienting \vell. \\

%%%%%%%%%%%%%%%%%%%%%%%%%%%%%%%%%%%%%%%%%%%%%%%%%%%%
%%%%%%%%%%%%%%%%%%       Figure SI1  (Figure 7)       %%%%%%%%%%%%%%%%%%%%%%%%%%%%%%%%%%%%%%%%%%%%%%%%%%%%%%%%%%%%%%%%%%%%%%%%
\begin{figure}
\centerline{\includegraphics[height=0.23\textheight]{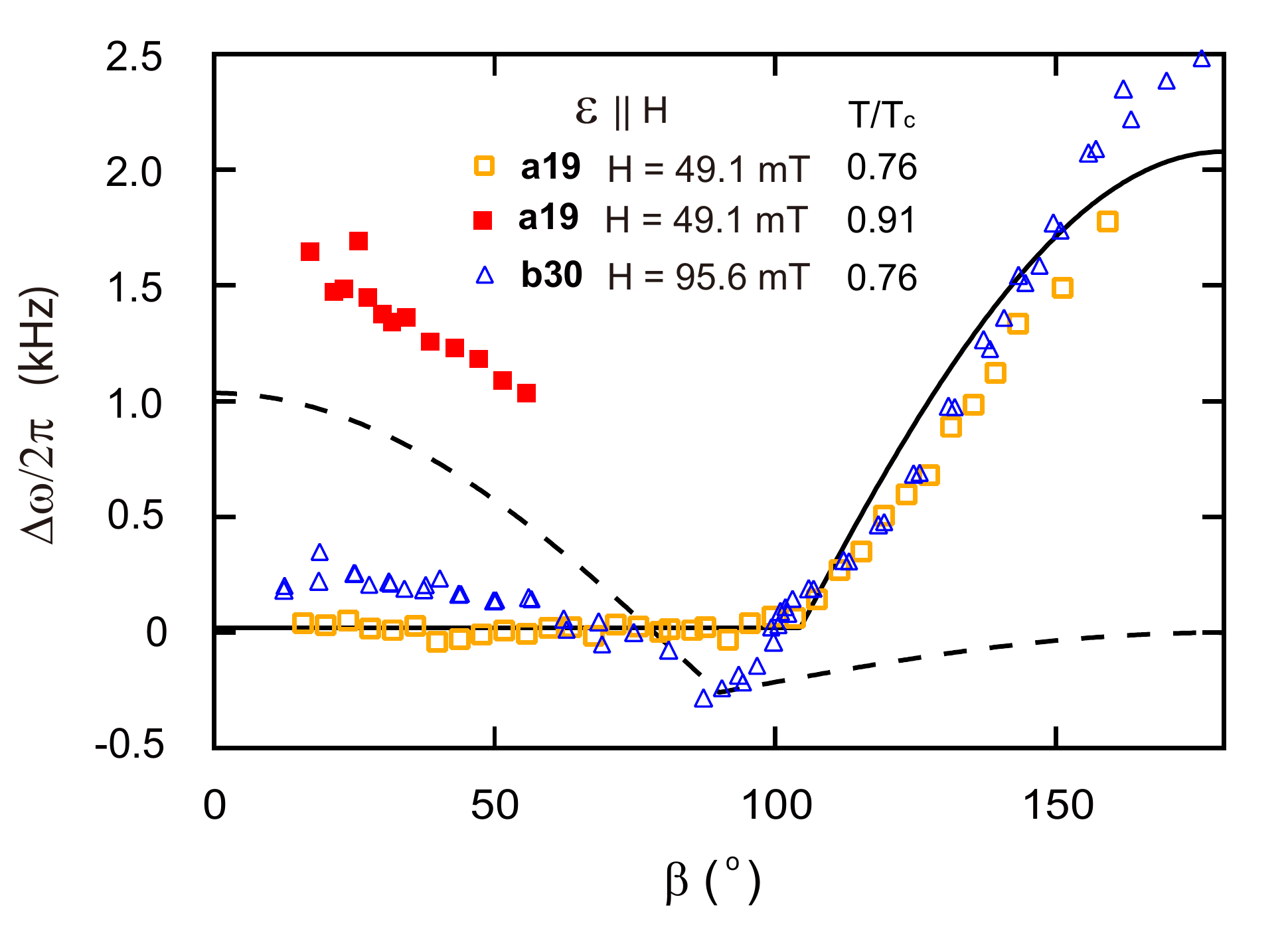}}
\caption{\label{figSI1} (Color online). Frequency shift versus tip angle for sample \aab\ and \bbc\ in the distorted B-phase. The magnitude of the shifts have been adjusted to a common field of $H = 95.6$ mT, a common temperature of $T/T_{ca} = 0.76$. The tip angle behavior of the isotropic sample $\mathbf{a0}$ is represented by the solid black curve  for the Brinkman-Smith-mode and the dashed curve for  the texture with \vell $\perp$ \vfield, Ref.~\cite{Pol.11}. The tip angle behavior of the anisotropic samples \blue{matches} well with the black solid curve below the textural cross-over, whereas above the cross-over it displays larger frequency \blue{shifts} than the maximum frequency \blue{shifts} observed in the isotropic sample \aaa. }
\end{figure}

%%%%%%%%%%%%%%%%%%%%%%%%%%%%%%%%%%%%%%%%%%%%%%%%%%%%
%%%%%%%%%%%%%%%%%%       Figure SI2   (Figure 8) %%%%%%%%%%%%%%%%%%%
%%%%%%%%%%%%%%%%%%%%%%%%%%%%%%%%%%%%%%%%%%%%%%%%%%%%

\begin{figure}
\centerline{\includegraphics[height=0.23\textheight]{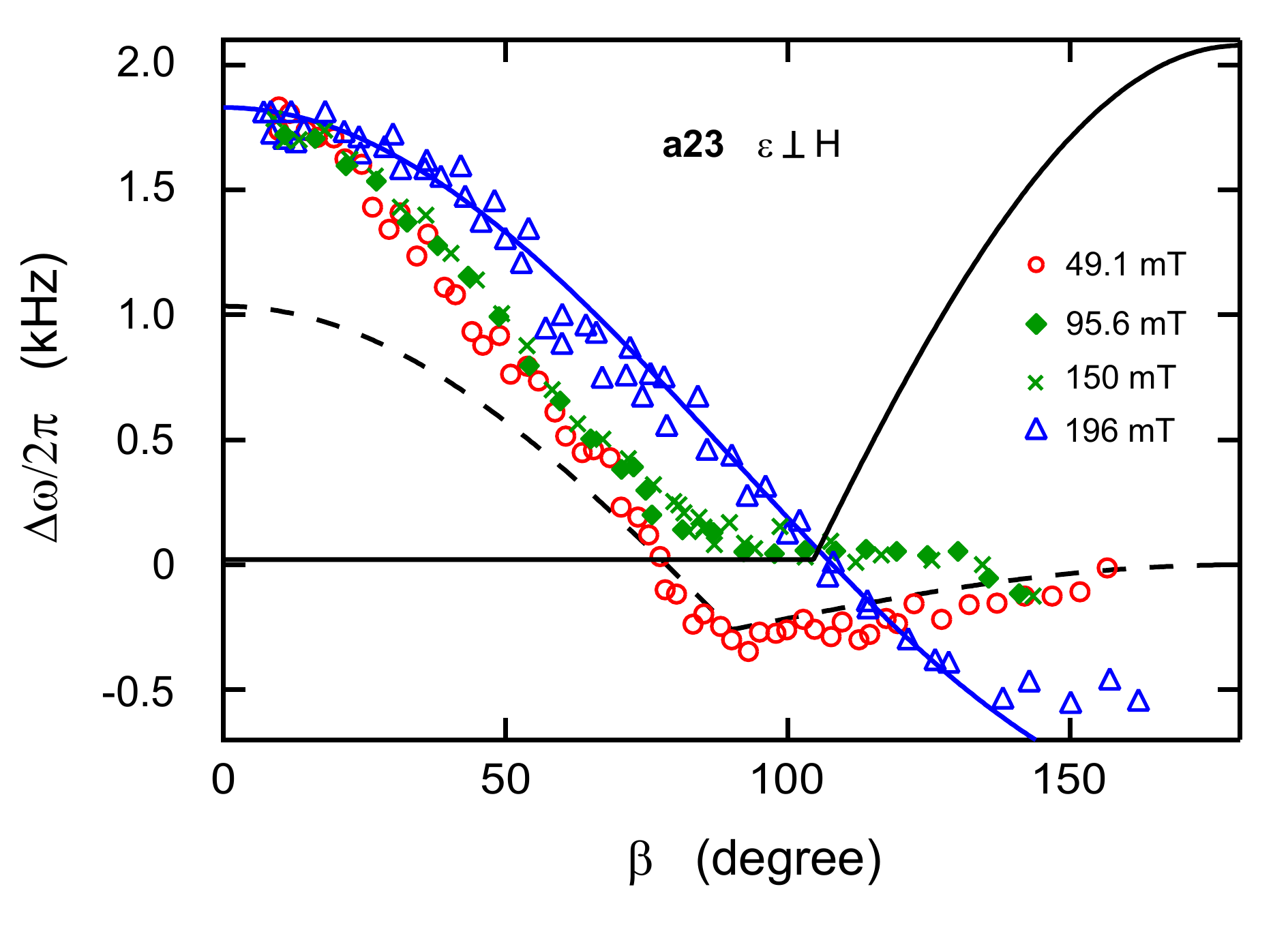}}
\caption{\label{figSI2} (Color online). Magnetic field dependence of the tip angle measurements for $\mathbf{a23}$ in the distorted B-phase at $T/T_c=0.46$. The magnitude of the shifts have been adjusted to a common temperature of $T/T_{ca} = 0.78$ and a common field of $H = 95.6$ mT. The solid black curve is the tip angle behavior for the Brinkman-Smith-mode and the dashed curve for the \vell $ \perp \bm{H}$-mode  of the isotropic sample $\mathbf{a0}$, Ref.~\cite{Pol.11}. In the magnetic field range of our measurements, the distorted B-phase behavior deviates significantly from the shift observed in sample \aaa. The blue solid curve is a fit to the data at $H = 196$ mT (solid blue circles), using a simple cosine function. }
\end{figure}

\noindent
{\bf SI.4\,\,\, Tip angle dependence in the distorted B-phase}\\

To explore the details of the dipole interaction of the distorted B-phase, we compare the tip angle dependence of the frequency shift, scaled to a common field of $H = 95.6$ mT using Eq.~6,  to the B-phase in the isotropic aerogel, Ref.~\cite{Pol.11}.  In the latter case, as in the present work, there is a  crossover between two order parameter textures as a function of temperature each with characteristic spin dynamics that can be identified from NMR tip angle measurements.  For the isotropic aerogel the two modes of the spin dynamics, stabilized by these textures, are the Brinkman-Smith-mode, Ref.~\cite{Bri.75}, at high temperatures, and below the crossover temperature, the \elperph-mode induced by  influence of the walls on the orbital quantization axis, \vell.  For the negatively strained aerogels  with strain axis parallel to the magnetic field, \ve\ $\parallel\bm{H}$, the tip angle behavior in sample \aab\ (orange squares) and \bbc\ (blue triangles) for the Brinkman-Smith-mode matches well with that of the B-phase measured in the isotropic aerogel (black solid curve), Ref.~\cite{Pol.11}, showing that this mode is unaffected by  negative strain parallel to the magnetic field.  However, the scaled frequency shift dependence on $\beta$ for the \elperph-mode in sample \aab\ (red squares) is significantly increased.  This result for small $\beta$ was noted in Fig.~4.

On the other hand, when the strain axis is perpendicular to the magnetic field, \ve\ $\perp \bm{H}$, the tip angle behavior of  the scaled frequency shift deviates significantly from either mode as shown in Fig.~SI\ref{figSI2}. At small tip angle $\beta$, the scaled  shift measured in sample \aac, is independent of the magnetic field, and significantly larger than the maximum frequency shift measured in the B-phase of the isotropic sample \aaa\ (black dashed curve), Ref.~\cite{Pol.11}. This is consistent with the small $\beta$ measurement shown in Fig.~4.   At large tipping angle, a magnetic field dependence is observed.  At $H = 49.1$ mT, the large tip angle frequency shift displays a sharp minimum at $\beta = 90^{\circ}$ and small onset at $\beta > 90^{\circ}$, similar to the \vell\ $\perp$ \vfield\ texture of the isotropic aerogel (black dashed curve). With increasing magnetic field, these features in tip angle behavior evolve at $H = 95.6$ mT and completely disappear at $H = 196$ mT. The blue solid curve in Fig.~SI\ref{figSI2} is a numerical fit to the $H = 196$ mT data using a cosine function, demonstrating that at $H =196$ mT, the tip angle behavior, and consequently the dipole interaction, is markedly different from the isotropic state (black solid and dashed curves) where neither of the B-phase modes (Brinkman Smith or \vell\ $\perp$ \vfield-mode) have magnetic field dependence of their scaled frequency shifts, Ref.~\cite{Pol.11}.\\

\noindent
{\bf SI.5\,\,\, Mean-free path analysis from scattering theory}\\

It is noteworthy that  fitting the AB-phase diagram slopes using the homogeneous isotropic scattering model, Ref.~\cite{Thu.98, Sau.03}, assuming that the distorted B-phase has the symmetry of the isotropic state, gives quasiparticle mean free paths, $\lambda$, from $g_{BA}/H^2_0$ to be $\lambda_{\mathbf{a0}} =210\,\mathrm{nm}, \lambda_{\mathbf{a19}} =250\,\mathrm{nm}, \lambda_{\mathbf{b12}} =215\,\mathrm{nm}, \lambda_{\mathbf{b20}} =240\,\mathrm{nm}$ and $\lambda_{\mathbf{b30}} =270\,\mathrm{nm}$.

\end{document}